# FRACTAL DIMENSIONS IN SWITCHING KINETICS OF FERROELECTRICS

J. F. Scott

Centre for Ferroics, Earth Sciences Department

University of Cambridge, Cambridge CB2 3EQ, U. K.

**Abstract**

Early work by the author with Prof. Ishibashi [Scott et al., J. Appl. Phys. **64**, 787 (1988)] showed that switching kinetics in ferroelectrics satisfy a constraint on current transients compatible with $d = 2.5$ dimensionality. At that time with no direct observations of the domains, it was not possible to conclude whether this was a true Hausdorff dimension or a numerical artefact caused by an approximation in the theory (which ignored the dependence of domain wall velocity upon domain diameter). Recent data suggest that the switching dimensionality is truly fractal with $d = 2.5$. The critical exponent $\beta$ characterizing the order parameter $P(T)$ can be written as a continuous function of dimension $d$ as $\beta(d) = [\nu(d)/2]\,[d+\eta(d)-2]$, which is exact within hyperscaling; here $\nu$ and $\eta$ are the exponents characterizing the pair correlation function $G(r,T)$ and the structure factor $S(q,T)$. For $d=2.5$ the estimate is that $\beta$ is approximately ¼.

**I. Introduction**

Twenty years ago the author used current transients $i(t) = dD/dt$ to determine details about switching kinetics in ferroelectric thin films; here $t$ is time and $D$ is the displacement vector, approximately equal to polarization $P$ in a ferroelectric. The initial study [1] used potassium nitrate, $KNO_3$, but similar studies of lead zirconate titanate PZT followed shortly thereafter [2]. Later work by Shur [3] extended this particular modelling technique. The theory used for this data-fitting was developed originally by Ishibashi and Takagi [4,5], and is based upon the analogous problem of grain growth in crystallization due to Avrami [6]. The theory is falsifiable in a nice way by the fact that it affords constraints that test the self-consistency of the model; in particular the dimensionless ratio

$$i(p)\, t(p)/P_s = \Lambda \qquad (1.)$$

must be satisfied, where i(p) is the maximum "peak" transient current value, which occurs at time t(p); $P_s$, the spontaneous polarization; and $\Lambda$, a dimensionless constant known independently [4,5] for each value of dimensionality d.

For well annealed PZT the value of d extracted from the shape of the i(t) data [2]

$$I(t) = (2P_s \: A \: d \: /t_o)(t/t_o)^{d-1} \: \exp[-(t/t_o)^d] \quad (2.)$$

yielded d = 2.55±0.25, for which the theory [4,5] predicted $\Lambda$ = 1.65. Here $P_s$ is spontaneous polarization. By comparison, the data inserted into Eq.(1.) gave $\Lambda$ = 1.64 ± 0.23. Thus the data were compatible with the theory (Eqs. 1 and 2) and implied a fractal dimension d = 2.5. Despite this good agreement, it was not justifiable to conclude that this required a true Hausdorff fractal dimension of 5/2, because other approximations inherent in the model could introduce good fits to parameters whose numerical values were artefacts of the model and its approximations; in particular, the theory [4,5] neglects the dependence of domain wall velocity v(r) upon the radius r of the nucleating domain during its initial spherical phase, and the fitting exponents are also affected by finite size of latent nuclei of reverse-polarized nano-domains.. The former dependence is given approximately by v(r) = a/r, where a is a constant; this correction was later shown to be important in the nucleation study of Ganpule et al. [7]. Thus, the possibility of fractal dimensionality lay dormant for a few years in ferroelectric switching studies.

This question was resurrected nicely by Paruch et al. [8], who inferred d = 2.5 dimensionality in the switching kinetics of PZT. This arises because the domain walls are not planar, and the roughening of the walls corresponds to a domain of Hausdorff dimension 5/2. And more recently the present author has given [9] tables of sets of possible critical exponents for d=2.5 that satisfy all known scaling and hyperscaling relationships, together with a fuller discussion of the possibility of d=2.5 Hausdorff dimensions. Kalinin has very recently observed fractal dimensions in his direct AFM measurement on nucleation of domains in ferroelectric switching processes [10]. Others have considered [11] that d might be 3/2 in ferroelectrics (Ref. [11] gives 1.0 < d < 1.7), but that does not seem plausible in a system which is three-dimensional in bulk [9], where d must be >2.0 [12].

## II. Predictions

For all temperatures reasonably far from the ferroelectric transition temperature, the rms fluctuations in polarization are much less than its mean value; in this situation the system can be described as "mean field," which is characteristic of the early models of Weiss for magnetism [13] or Van der Waals for fluids [14]. But very near the phase transition temperature in ferroelectrics, especially near second-order transitions, it is expected that fluctuations in the order parameter P(T) become large such that $<P^2> >> <P>^2$. Under such circumstances the system is called "critical", and the dynamics transcend the structural details of the material. Such phenomena are well known in fluids and magnets, but not in ferroelectrics or superconductors. In the latter systems the interactions are quite long-range in real space (no matter how close the Cooper pairs are shown in naïve textbooks!), and according to the Ginzburg Criterion (1960), the temperature range over which "critical" fluctuations are dominant varies as $L^{-6}$, where L is the interaction length in the system. For ferroelectrics, L is typically large (microns?), and the fluctuation region $\Delta T$ near $T_0$ is probably < 0.001 K in width. Parenthetically we note that the so-called "Ginzburg Criterion" is misnamed; it was actually published by A. P. Levanyuk (a student of Ginzburg) as sole author in 1959, a year before Ginzburg republished it. So we will henceforth call it the "Levanyuk-Ginzburg Criterion" (although it probably *should* be the "Levanyuk Criterion").

**III. Critical phenomena in ferroelectrics – a quick primer**

There are four common "critical" exponents describing static phenomena near a ferroelectric phase transition temperature $T_0$: $\alpha$ describes the divergence of specific heat at constant field $C_E(T) = a\,\tau^{-\alpha}$, where $\tau$ is reduced temperature $(T-T_c)/T_c$; $T_c$ is the Curie temperature, strictly defined as the temperature at which the inverse electric susceptibility extrapolates to zero from temperatures above $T_0$, and equals $T_0$ for continuous (second-order) transitions; $\beta$ describes the temperature dependence of the polarization $P(\tau) = b\,\tau^{\beta}$; $\gamma$ describes the divergence in isothermal electric susceptibility $\chi(\tau) = c\,\tau^{-\gamma}$; and $\delta$ describes the dependence of polarization $P(E)$ upon applied field E along the critical isotherm, $P(E) = e\,E^{1/\delta}$. There are similar exponents describing dynamical properties, such as that for ultrasonic attenuation divergence $\zeta$, thermal conductivity, or Landau-Placzek ratio. There are a dozen known thermodynamic relations which constrain these critical exponents, such as the Rushbrooke Equation $[\alpha + 2\beta + \gamma = 2]$, the Griffiths Equation $[\alpha + \beta(1+\delta) = 2]$, the Widom Equation $[\gamma = \beta(\delta-1)]$, etc. Generally, these are, from thermodynamics, inequalities; but the inequalities become exact equalities under the assumption of "static scaling." Scaling also implies that $\alpha = \alpha'$; $\gamma = \gamma'$; etc., where primes denote values below $T_0$ and unprimed, above $T_0$. It is useful to note that in mean field at continuous phase transitions, $\alpha = 0$ (step discontinuity or logarithmic divergence); $\beta = \frac{1}{2}$; $\gamma = 1$; and $\delta = 3$. In mean field at "tricritical" points (where a phase transition goes from second-order to first-order as some parameter – usually pressure – is changed), $\alpha = \frac{1}{2}$; $\beta = \frac{1}{4}$; $\gamma = 1$; and $\delta = 5$.

In addition to the scaling equations and the static scaling "critical" exponents, there are additional relations called "hyperscaling". Hyperscaling equalities involve explicitly the dimensionality d of the system. This leads to categories such as the [2D] Ising model, the [3D] Heisenberg model, the "spherical" (infinite-dimensional) model, etc. Such models belong to "universality classes", in which the phase transition dynamics and exponents are the same, independent of whether the system is a magnet, ferroelectric, liquid crystal, superconductor, etc. Hyperscaling necessarily introduces two additional exponents, $\nu$ and $\eta$, which relate respectively to the correlation function $G(r,\tau)$ and the structure factor $S(q,\tau)$.

During the period of 1950s-1960s physicists found that almost every conceivable model had some physical realization in different magnetic structures, and a kind of sport developed in measuring carefully the various critical exponents to show that such descriptions were self consistent with the model predictions. It is not an exaggeration to say that a whole cottage industry developed for this task! Unfortunately during the 1970s many physicists tried to find such critical exponents and universality classes in ferroelectrics. Notable is the book by Bruce and Cowley [15]. In my opinion this was not a good idea (based upon the Levanyuk-Ginzburg Criterion). A few intrepid souls have continued this endeavour (e.g., Kleemann, Dec et al. [15]). The present author's opinion is that the existence of non-mean-field critical exponents in ferroelectrics is not well demonstrated experimentally, with the possible exception of ferroelectric liquid crystals.

Much of the mischief was caused by Nobel Prize-winner K. A. Mueller, who inferred $\beta = 0.332$ for the antiferrodistortive transition in strontium titanate [16]. His fit to his EPR data on $Fe^{+3}$ at $Ti^{+4}$ sites is unreliable for three reasons: (1) He fitted an arbitrary temperature region without reliable independent measurement of $T_0$; and $T_0$ and $\beta$ are highly correlated in the fit; (2) no other scientist has ever verified this result, and several have found that a mean-field result works much better (Kozlov et al. [17]; Hayward et al., [18]); and (3), it is unreasonable to deduce critical exponents from any measurements on defect sites, since defects are known to severely perturb the dynamics near phase transitions.[19-22] Similarly, the claim by Courtens [23] that birefringence in $SrTiO_3$ near $T_0 = 105$ K is also non-mean-field was disproved by Glazer et al. recently [24]. Finally, the claim of other non-mean-field exponents in this material by Cowley and Nelmes was shown later at Brookhaven to arise from defective surface regions and disappeared when such entrance/exit face scattering was masked off from the neutron scattering.

An equally important problem is the existence of changes in apparent exponents in ferroelectrics due to defects. First elucidated by Larkin and Khmelnitskii [19,20], this was developed into a detailed theory by Levanyuk and Sigov [21, 22], and independently, by Imry and

Wortis [21].  These have been experimentally measured unambiguously by Ryan et al. [22] and Nelmes et al. [25] via X-ray and neutron scattering.

**IV. Defect models**

The defect model shows that near $T_0$ in real ferroelectrics the various thermodynamic parameters can display exponents that are NOT due to fluctuations and hence not from a universality class; but they are very different from mean-field values.  These are listed in Table I below:

Table 1 Critical exponents in Levanyuk-Sigov defect theory

| $\alpha$ | $\beta$ | $\gamma$ | $\delta$ |
|---|---|---|---|
| 3/2 | Ca. 0.4 | 5/2 | 2 |

These predictions agree reasonably well with the experimental values in strontium barium niobate (SBN), a well-studied relaxor ferroelectric, where Kleemann et al. find [15] $\gamma = 1.85$-$2.10$ and $\delta = 1.53 \pm 0.15$, and Chen and Scott [26] find $\beta = 0.33 \pm 0.05$.

To make contact with our earlier section of the present paper dealing with fractal dimensions, Scott has found [9] several sets of numerical values for dimension $d = 2.5$ which satisfy as equalities all known scaling and hyperscaling equations.  Two plausible sets are given in Table 2.  Here the top row gives values that assume correlation exponent $\nu = 1$ as in [2D] Ising models; the lower row gives values compatible with $\nu = 2/3$ – as in [3D] Ising models.

Table 2 Critical exponents in $d = 2.5$ dimensions (two possible sets satisfying scaling)

| $\alpha$ | $\beta$ | $\gamma$ | $\delta$ |
|---|---|---|---|
| -1/2 | 1/4 | 2 | 9 |
| +1/3 | 1/3 | 1 | 4 |

It would be useful to do careful measurements of $\alpha$ in SBN to test these models.  Unfortunately the best specific heat data on SBN are those of Yakushkin [27], which show that the dependence fits no exponent and in fact is highly nonequilibrium (very different data for cooling and heating and relaxing with time at constant temperature).  This vividly confirms the comment of Levanyuk and Sigov that relaxors are usually not in thermal equilibrium but have residual fields.  Unfortunately Yakushkin fits his specific heat data $C(T)$ in SBN to a Schottky anomaly.  Schottky anomalies normally occur at ca. 1 K, not 300 K, and they have nothing whatsoever to do with phase transitions; so the fitting in [27] is difficult to justify.

It is only slightly speculative to incorporate the possible value of $\beta = 1/4$ for d=2.5 from Table 2 into Table 3 below, which shows how $\beta$ varies with dimensionality, including the d = infinity "spherical" model. The suggestion is that $\beta(d)$ probably varies continuously from 1/8 for d = 2 to ½ for infinite d.

Table 3 Critical exponent $\beta$ as a function of dimensionality for Ising-like systems

| Dimension d=2 | d = 2.5 | d = 3 | d = infinity |
|---|---|---|---|
| 1/8 | 1/4 | 5/16 | 1/2 |

Note that it is quite useful to express $\beta(d)$ in scaling theory as

$$\beta(d) = (\nu/2)(d+\eta-2) \tag{3a.}$$

This is a very convenient expression for $\beta(d)$, since for example d=2 gives $\beta = 1/8$; $\nu = 1$; and $\eta = ¼$; which satisfy (3a.) And d=3 gives $\beta = 5/16 = 0.313$; $\nu = 0.638$; and $\eta = 0.041$; which also satisfy (3a.). But is not as handy as it might seem at first, because $\nu$ and $\eta$ vary very rapidly with d. Therefore it is wiser to write Eq.(3a.) as

$$\beta(d) = [\nu(d)/2][d+\eta(d)-2]. \tag{3b.}$$

If we make a crude linear interpolation between the known values of $\nu$ and $\eta$ for d = 2 and d =3, we estimate $\nu = 0.8$ and $\eta = 0.15$ for d = 2.5. Substituting these values into Eq.(3b.) yields $\beta = 0.22$, which is in good accord with our value 0.25 in Tables 2 and 3 which were obtained independently.

**V. Summary**

Based upon the recent work of Paruch et al. and Kalinin, it appears that there is a likelihood that the d = 2.5 fractal dimensionality inferred some years ago for switching kinetics in potassium nitrate and PZT corresponds to a real Hausdorff fractal dimensionality. The author has recently given sets of plausible critical exponents for d=2.5, but the test of such models for real ferroelectrics is complicated by the possibility of defect-dominated dynamics.

Acknowledgement: I thank P. Zubko for helpful discussions.